\documentstyle[prl,aps,preprint]{revtex}

\def\vec#1{\bbox{#1}}
\def\vecset#1{\tilde{\vec{#1}}}
\def\tr{\mbox{\rm tr}}

\def\opsigmaN{\hat{\vec\sigma}\cdot\vec a_1\otimes\cdots\otimes
              \hat{\vec\sigma}\cdot\vec a_N}

\def\tr{\mbox{\rm tr}}

\def\nN{\vec n_1,\ldots,\vec n_N}
\def\aN{\vec a_1,\ldots,\vec a_N}
\def\amN{(\vec a_1\cdot\vec m_1)\cdots(\vec a_N\cdot\vec m_N)}
\def\projN{|\vec n_1\rangle\langle\vec n_1|\otimes\cdots\otimes
           |\vec n_N\rangle\langle\vec n_N|}
\def\tildeprojN{|\vecset n\rangle\langle\vecset n|}
\def\LambdaN{\Lambda_1,\ldots,\Lambda_N}

\begin{document}
\tighten

\title{Local Realistic Model for the Dynamics of \\
Bulk-Ensemble NMR Information Processing}

\author{N.~C. Menicucci and Carlton M.~Caves}

\address{Department of Physics and Astronomy,
University of New Mexico, Albuquerque, NM 87131-1156}

\date{November 19, 2001}

\maketitle

\begin{abstract}
We construct a local realistic hidden-variable model that
describes the states and dynamics of bulk-ensemble NMR information
processing up to about 12 nuclear spins.  The existence of such a
model rules out violation of any Bell inequality, temporal or
otherwise, in present high-temperature, liquid-state NMR
experiments.  The model does not provide an efficient description
in that the number of hidden variables grows exponentially with
the number of nuclear spins.
\end{abstract}

\vspace{24pt}

High-temperature, liquid-state nuclear magnetic resonance (NMR)
provides a testing ground for the new ideas for information
processing that are being developed in quantum information science
\cite{Nielsen2000a}.  The fundamental information-processing
elements used in NMR are two-level nuclear spins, called {\it
qubits}, which are bound together in a single molecule.  A liquid
NMR sample contains a macroscopic number of molecules, each of
which functions as an independent information-processing unit. The
molecules are initially in thermal equilibrium at high enough
temperature that the nuclear spins are only weakly polarized along
the direction of a strong magnetic field.  NMR techniques cannot
control the quantum states of individual molecules, and the
measurements performed in NMR detect the average magnetization of
the entire sample.  For these reasons the use of high-temperature,
liquid-state NMR to emulate quantum computation is called {\it
bulk-ensemble quantum computation}.

The original proposals \cite{Cory1997a,Gershenfeld1997a} for
quantum information processing using NMR were greeted with
enthusiasm tempered by skepticism.  The enthusiasm led to a
remarkable series of experiments in which NMR techniques have been
used to implement the operations for a variety of
quantum-information-processing jobs involving up to seven qubits
(for reviews of NMR information
processing, see Refs.~\cite{Havel2000a,Cory2000a,JJones2000a,%
JJones2001a,JJones2001b,Vandersypen2001a}).  The persistent
skepticism has to do with questions about the ``quantumness'' of
NMR information processing.  Initially based on doubt that the
highly mixed states used in NMR could be used to achieve genuinely
quantum-mechanical effects, these questions were made concrete by
the realization that all the quantum states accessed in present
experiments are {\it unentangled\/} \cite{Braunstein1999a}.
Entanglement is often thought to be an essential feature of
quantum computation \cite{Jozsa1998c,Ekert1998a}.  Arguments for
an essential ``quantumness'' in NMR information processing are
presented in Ref.~\cite{Laflamme2001a}, and an entirely different
method for characterizing the ``quantumness'' of NMR is developed
by Poulin \cite{Poulin2001a}.

The absence of entanglement in present NMR experiments means that
the statistics of measurements made at any time during the
experiments can be understood in terms of a {\it local
realistic\/} hidden-variable model \cite{Peres1993a} in which each
spin has objective properties that determine the results of the
measurements. In a local realistic model, the correlations
observed in experiments can be attributed to classical
correlations between realistic properties of the component qubits.
The ability to describe the correlations observed in NMR
experiments in terms of classical correlations between the qubits
casts doubt on the ``quantumness'' of the experiments.

Modelling the statistics of the measurements made in NMR
experiments is not sufficient for understanding the experiments.
One must also be able to model the dynamics of the nuclear spins.
NMR experimenters can implement with high accuracy any unitary
operation, including the nonfactorizable unitary
operations---those that cannot be written as a product of
unitaries for each qubit---that produce entanglement when applied
to pure quantum states.  Previous attempts \cite{Schack1999b} to
devise a local realistic description of the dynamics were only
partially successful in that they did not provide a local
realistic description of the changes produced by nonfactorizable
unitaries which reproduced all the predictions of quantum
mechanics.  This left open the possibility that one might not be
able to describe the correlations observed in successive
measurements separated by nonfactorizable unitary operations in
terms of local realistic properties and thus that present NMR
experiments might violate {\it temporal\/} Bell inequalities
\cite{temporal} for successive measurements.

In this Letter we report a local realistic hidden-variable (LRHV)
model for the states and dynamics of bulk-ensemble NMR information
processing up to about 12 qubits.  The existence of such a model
rules out violation of any Bell-type inequality in present NMR
experiments.  This conclusion applies only to the bulk-ensemble
model of information processing realized in present
high-temperature, liquid-state NMR experiments; it does not apply
to NMR methods based on distilling a pure state from a thermal
state \cite{Cory2000a,LJSchulman1999a,Knill1998d}.  Our model is
not satisfactory from the point of view developed by Schack and
Caves \cite{Schack1999b} because the number of hidden variables
scales exponentially with the number of qubits.

All NMR quantum computing experiments performed so far work in the
following way.  The state of each molecule, consisting of $N$
active spin-$1\over2$ nuclei, is described by a density operator
\begin{equation}
  \hat\rho = (1-\epsilon)\hat 1/2^N + \epsilon\hat\rho_1 \;,
  \label{eq:rhoepsilon}
\end{equation}
which is a mixture of the desired state of the quantum computer,
$\hat\rho_1$, with the maximally mixed state for $N$ qubits, $\hat
1/2^N$, $\hat 1$ being the unit operator. When $\hat\rho_1$ is a
pure state, $\hat\rho$ is called a {\it pseudopure state\/}
\cite{Cory1997a}.

The molecules in an NMR sample begin in thermal equilibrium, with
a weak polarization $\alpha=h\nu/2kT\sim2\times10^{-5}$ at room
temperature, where $\nu\sim300\,$MHz is the average resonant
frequency of the active spins in the strong longitudinal magnetic
field. The first step in NMR information processing is to
transform the molecules from equilibrium to a pseudopure state
\cite{Gershenfeld1997a,pseudopure}. A consequence of pseudopure
state synthesis is that the mixing parameter scales like
$\epsilon=\alpha N/2^N$.

After synthesis of the desired initial state, the computation
begins. The unitary operations required for the computation can be
constructed from sequences of radio-frequency pulses alternating
with periods of continuous evolution under the nuclear-spin
Hamiltonian \cite{Havel2000a,Cory2000a,JJones2000a,%
JJones2001a,JJones2001b,Vandersypen2001a}.  A unitary operator
$\hat U$ takes an input state $\hat\rho$ to an output state
\begin{equation}
\hat U\hat\rho\,\hat U^\dagger =(1-\epsilon)\hat 1/2^N + \epsilon
\hat U\hat\rho_1 \hat U^\dagger\;. \label{eq:paradigm}
\end{equation}
The maximally mixed state is unaffected by the unitary
transformation.  The output state retains the
form~(\ref{eq:rhoepsilon}) with the same value of $\epsilon$,
and---this is the essence of the bulk-ensemble paradigm for
quantum computation---$\hat\rho_1$ undergoes the desired unitary
transformation.

The computation completed, the last step is to read out the
answer.  By applying radio-frequency pulses and then measuring the
transverse magnetization of the sample, an NMR experimenter can
determine the expectation value of any product of spin components,
one for each qubit \cite{Havel2000a,Cory2000a,JJones2000a,%
JJones2001a,JJones2001b,Vandersypen2001a}.  These expectation
values have the form
\begin{eqnarray}
C(\vecset a)&\equiv&C(\aN)\nonumber\\
&=&\tr(\hat\rho\,\opsigmaN)\nonumber\\
&=&\epsilon\,\tr(\hat\rho_1\,\opsigmaN) \;. \label{eq:expect}
\end{eqnarray}
Here and throughout, a tilde over a quantity denotes a collection of
$N$ such quantities, one for each spin.  In Eq.~(\ref{eq:expect}) the
tensor product includes one operator for each spin; the vector
operator $\hat{\vec\sigma}\equiv\hat 1\vec e_0+\hat\sigma_x\vec e_x+
\hat\sigma_y\vec e_y+\hat\sigma_z\vec e_z$, where $\hat\sigma_x$,
$\hat\sigma_y$, and $\hat\sigma_z$ are the Pauli operators; and $\vec
a_r$ is either a spatial unit vector, in which case
$\hat{\vec\sigma}\cdot\vec a_r$ is the component of spin $r$ along
the direction $\vec a_r$, or the unit vector $\vec e_0$ in the
``zero'' direction, in which case spin $r$ does not contribute to the
expectation value.  The last equality in Eq.~(\ref{eq:expect})
assumes at least one of the vectors $\vec a_r$ is a spatial
direction.

The expectation values~(\ref{eq:expect}) express the correlations
between spin components of different spins.  The maximally mixed
state does not contribute to the correlation coefficients, which are
determined by the state $\hat\rho_1$ that undergoes the desired
evolution.  The mixing parameter $\epsilon$ measures the strength of
the magnetization signal.  The scaling $\epsilon=\alpha N/2^N$ that
comes from pseudopure state synthesis thus leads to an in-principle
demand for an exponentially increasing number of molecules as the
number of qubits increases \cite{Cory1997a,Warren1997a}, implying
that bulk-ensemble quantum computation is not suitable for
large-scale quantum computation.

Before constructing our LRHV model, we recall that any $N$-qubit
density operator $\hat\tau$ has an associated quasidistribution
\cite{Schack2000a}
\begin{equation}
    w_{\hat\tau}(\vecset n)\equiv
    \tr\Bigl(\hat\tau\,\hat Q(\vecset n)\Bigr)\;,
    \label{eq:wrho}
\end{equation}
where the vectors in the set $\vecset n\equiv(\nN)$ are spatial
unit vectors and
\begin{equation}
    \hat Q(\vecset n) \equiv {1\over{\cal N}^N}
    (\hat 1+3\vec n_1 \cdot \hat{\vec\sigma})
    \otimes \cdots \otimes
    (\hat 1+3\vec n_N\cdot\hat{\vec\sigma})
    \;.
\label{eq:Q}
\end{equation}
For each spin, the unit vector $\vec n$ can point in ${\cal N}$
different directions satisfying
\begin{eqnarray}
0&=&\sum_{\vec n} n_j\;,
\label{eq:0nj}\\
{1\over3}\delta_{jk}&=&{1\over{\cal N}}\sum_{\vec n}n_jn_k\;,
\label{eq:isotropic}
\end{eqnarray}
where the sums are over the possible directions and the subscripts
indicate spatial components of $\vec n$.
Condition~(\ref{eq:isotropic}) means that the vectors
$\sqrt{3/{\cal N}}\vec n$ form a resolution of the 3-dimensional
unit tensor; condition~(\ref{eq:0nj}) places an additional
constraint on the placement of the vectors.   The vertices of a
tetrahedron give the minimum number, ${\cal N}=4$, of possible
directions.  The six vectors along the cardinal directions make up
another simple possibility.

The density operator is then given by \cite{Schack2000a}
\begin{equation}
\hat\tau=\sum_{\vecset n}w_{\hat\tau}(\vecset n)\tildeprojN\;,
\label{eq:doexpansion}
\end{equation}
where $\tildeprojN\equiv\projN$ and $|\vec n\rangle\langle\vec
n|={1\over2}(\hat1+\hat{\vec\sigma}\cdot\vec n)$ is the $+1$
eigenstate of $\hat{\vec\sigma}\cdot\vec n$.  In terms of
$w_{\hat\tau}(\vecset n)$, the correlation
coefficients~(\ref{eq:expect}) take the form
\begin{equation}
C(\vecset a)= \sum_{\vecset n}w_{\hat\tau}(\vecset n)\amN\;,
\label{eq:Ca}
\end{equation}
where $\vec m_r\equiv\vec n_r+\vec e_0$.

Under a unitary transformation the quasidistribution evolves
according to
\begin{equation}
w_{\hat U\hat\tau\hat U^\dag}({\vecset n}')= \sum_{\vecset
n}T^{\hat U}_{{\vecset n}'\vecset n}w_{\hat\tau}(\vecset n)\;.
\label{eq:t}
\end{equation}
Here the transformation matrix $T^{\hat U}$ has matrix elements
\begin{equation}
T^{\hat U}_{{\vecset n}'\vecset n}\equiv
\langle\vecset n|\hat U^\dag\hat Q({\vecset n}')\hat U|\vecset n\rangle
= w_{\hat U|\vecset n\rangle\langle\vecset n|\hat U^\dag}({\vecset n}')\;,
\label{eq:tmatrix}
\end{equation}
given by the quasidistribution for $\hat U|\vecset n\rangle$.

A {\it separable\/} density operator is one that has an ensemble
decomposition in terms of product states.  Such a state has no
entanglement.  If the quasidistribution $w_{\hat\rho}(\vecset n)$
is everywhere nonnegative, then $\hat\rho$ is definitely
separable, and the statistics of all measurements can be
understood in terms of classical tops whose probability to point
in the directions $\vecset n$ is $w_{\hat\rho}(\vecset n)$.

For any density operator, the quasidistribution satisfies
$w_{\hat\rho}(\vecset n)\ge\hbox{}$[minimum eigenvalue of $\hat
Q(\vecset n)]=(-2)4^{N-1}/{\cal N}^N=-2^{2N-1}/{\cal N}^N$
\cite{Schack1999b,Schack2000a}. Thus for density operators of the
form~(\ref{eq:rhoepsilon}), the quasidistribution is everywhere
nonnegative if \cite{Braunstein1999a,Schack2000a}
\begin{equation}
\epsilon\le{1\over1+2^{2N-1}}\equiv\eta\;.
\end{equation}
Such states are {\it unentangleable\/} by any unitary
transformation. For the polarization $\alpha\sim2\times10^{-5}$ of
present NMR experiments, all states up to about 12 qubits are
unentangleable.  It is known \cite{Duer1999a} that entangled
states of the form~(\ref{eq:rhoepsilon}) exist for
$\epsilon>(1+2^{N-1})^{-1}\equiv\eta'$, i.e., $N\agt2/\alpha$, but
whether there are entangled states for $\eta<\epsilon\le\eta'$ is
an open question.

We turn now to constructing a LRHV model for unentangleable states,
i.e., for $\epsilon\le\eta$.  A straightforward model regards the
directions $\vecset n$ as hidden spin directions that determine the
results of measurements stochastically.  The problem with this simple
model is that the change in the probability distribution
$w_{\hat\rho}(\vecset n)$ as a consequence of a unitary
transformation should be described by transition probabilities, which
give the probability to go from initial directions $\vecset n$ to
final directions $\vecset n'$.  The matrix
elements~(\ref{eq:tmatrix}) provide candidate transition
probabilities, but they cannot be interpreted as transition
probabilities because they generally take on negative values. Schack
and Caves \cite{Schack1999b} derived nonnegative transition
probabilities from these matrix elements, but the dynamics described
by these transition probabilities did not duplicate the predictions
of quantum mechanics.

To include the unitary dynamics in the local realistic
description, we construct a deterministic LRHV model that includes
the probabilities $w_{\hat\rho}(\vecset n)$ in the set of hidden
variables.  A quasidistribution $w_{\hat\tau}(\vecset n)$ can be
regarded as a vector $\bar w_{\hat\tau}$ with ${\cal N}^N$
components labelled by the directions $\vecset n$.  An arbitrary
vector in this space, denoted by $\bar w$, has components
$w(\vecset n)$.  We are only interested in normalized vectors,
i.e., $\sum_{\vecset n}w(\vecset n)=1$.

The hidden variables in our model are a vector $\bar w$, a set of
spin directions $\vecset n$, and a set of real variables
$\tilde\Lambda\equiv(\LambdaN)$ such that $-1\le\Lambda_r\le1$,
$r=1,\ldots,N$.  The hidden variables are denoted collectively by
$\lambda=(\bar w, \vecset n, \tilde\Lambda)$.  The probability
density for $\lambda$ is
\begin{equation}
P(\lambda)={1\over2^N}\delta(\bar w - \bar w_{\hat\rho})w(\vecset
n)\;,
\end{equation}
indicating that the variables $\tilde\Lambda$ are distributed
randomly, the hidden vector $\bar w$ has a definite value given by
$\bar w_{\hat\rho}$, and the probabilities for the hidden spin
directions are determined by the hidden vector $\bar w$ and,
hence, are written as $w(\vecset n)$.  One can think of the hidden
vector $\bar w$ as a set of parameters that weight a ``roulette
wheel,'' so that when the wheel is ``spun'' to generate hidden
spin directions, the probability for directions $\vecset n$ is
$w(\vecset n)$.  Notice that the model requires that all components
of $\bar w_{\hat\rho}$ be nonnegative, since these components
become probabilities for the spin directions.

Measurement results are governed by functions $A_r(\vec
a_r,\lambda)$, $r=1,\ldots,N$.  The value of $A_r(\vec
a_r,\lambda)$, either $\pm1$, determines the result of a
projective measurement of the component of spin~$r$ along spatial
unit vector $\vec a_r$.  The model is {\it realistic\/} because
the results of spin-component measurements are determined by the
hidden variables, and it is {\it local\/} because the result of a
measurement of the component of spin~$r$ along direction $\vec
a_r$ depends only on $\vec a_r$ and the hidden variables, not on
measurement directions for other spins.

We choose the spin-component functions to be
\begin{equation}
A_r(\vec a_r,\lambda)=A_r(\vec a_r,\Lambda_r,\vec n_r)
=\cases{+1,&if $\Lambda_r\ge-\vec a_r\cdot\vec m_r$,\cr
       -1,&if $\Lambda_r<-\vec a_r\cdot\vec m_r$.}
\label{eq:Ar}
\end{equation}
By using $\vec m_r\equiv\vec n_r+\vec e_0$, we can employ this
function in cases where spin~$r$ is not involved in a measurement,
i.e., when $\vec a_r=\vec e_0$, which gives $A_r(\vec
e_0,\lambda)=1$. The correlation coefficients predicted by the
LRHV model,
\begin{eqnarray}
C_{\rm LRHV}(\vecset a) &=&\int d\lambda\,P(\lambda) A_1(\vec
a_1,\lambda)\cdots A_N(\vec a_N,\lambda)
\nonumber \\
&=&\sum_{\vecset n}\int d\bar w\, \delta(\bar w - \bar
w_{\hat\rho})w(\vecset n)
\nonumber \\
&\mbox{}&\qquad\quad\times\prod_{r=1}^N {1\over 2} \int_{-1}^{+1}
d\Lambda_r\, A_r(\vec a_r,\Lambda_r,\vec n_r)
\nonumber \\
&=&\sum_{\vecset n}w_{\hat\rho}(\vecset n)\amN\;,
\end{eqnarray}
duplicate the predictions of quantum mechanics.

Unitary dynamics fits easily into this model.  For any unitary
transformation $\hat U$, the hidden variables for each molecule are
updated in the following way: the variables $\tilde\Lambda$ are
chosen randomly, the hidden vector $\bar w$ is updated
deterministically using the transition matrix as in Eq.~(\ref{eq:t}),
and a new set of hidden spin directions is obtained by spinning a
roulette wheel weighted by the new hidden vector, $\bar w_{\hat
U\hat\rho\hat U^\dagger}$.  By treating the components of the hidden
vector not as probabilities, but rather as parameters that are used
to generate hidden spin directions stochastically, we avoid the need
for a nonnegative transition matrix.

This description of the updating applies to a discrete unitary
transformation, but it is easy to generalize it to a
quasicontinuous hidden-variable dynamics: each molecule updates
randomly in the fashion just described; i.e., each has a
probability $\gamma dt$ to update within each time interval $dt$.
The only requirement on this quasicontinuous dynamics is that the
mean time $\gamma^{-1}$ between updates be large compared to the
precession time $\nu^{-1}$ of the nuclear spins in the strong
magnetic field, but shorter than the duration of the
radio-frequency pulses that are used to produce the desired
dynamics.

It is trivial to generalize the LRHV model presented here to
nonunitary evolutions, since these evolutions are, like unitary
transformations, linear in the density operator. It is likely that
the model could be extended to include all dynamics that accesses
only separable states, i.e., states that have an expansion like
Eq.~(\ref{eq:doexpansion}), but with more general nonnegative
quasidistributions than the canonical form~(\ref{eq:wrho}).

The LRHV model developed here achieves our purpose of determining
whether present NMR experiments can violate Bell inequalities.
Bell inequalities are founded on two assumptions: the assumption
that systems have objective properties and a ``no-disturbance''
assumption that asserts that the relevant measurements report
faithfully the values of these properties.  Standard Bell
inequalities \cite{Peres1993a} justify the no-disturbance
assumption from locality---measurements here cannot affect
properties there.  Temporal Bell inequalities \cite{temporal},
which involve successive measurements on a system, have a tougher
time justifying the no-disturbance condition.  Our LRHV model for
NMR experiments describes the statistics of all measurements in
terms of evolving classical correlations between realistic
properties of the constituent nuclear spins.  The conclusion is
that NMR experiments up to about 12 qubits cannot violate any Bell
inequality, temporal or otherwise.

Our purpose achieved, we acknowledge that our LRHV model is
terribly contrived.  It succeeds in giving a local realistic
description of the dynamics by the brute force device of including
an encoding of the entire density operator among the hidden
variables.  As a result, it requires an exponentially increasing
number of hidden variables, $\sim4^N$ in the most efficient
version of the model.  It leaves open the possibility that
\cite{Schack1999b} the ``quantumness'' of NMR information
processing lies in the ability to implement nonfactorizable
unitaries that do not have an efficient local realistic
description.

This work was partly supported by the National Security Agency
(NSA) and the Advanced Research and Development Activity (ARDA)
under Army Research Office (ARO) Contract No.~DAAD19-01-1-0648.
CMC received support from National Science Foundation Contract
No.~PHY99-07949 at the Institute for Theoretical Physics of the
University of California, Santa Barbara.

\end{document}